\documentclass[pra,twocolumn,superscriptaddress,showpacs,preprintnumbers,amsmath,amssymb]{revtex4}
\usepackage{graphicx}% Include figure files
\usepackage{dcolumn}% Alig table columns on decimal point
\usepackage{bm}% bold math
\usepackage{enumerate}
\usepackage{float}
\begin{document}

\bibliographystyle{apsrev} % Choose Phys. Rev. style for bibliography, Rev.4
%\preprint{APS/123-MW}

\title{Preserving coherent spin and squeezed spin states of a spin-1 Bose-Einstein condensate with rotary echoes}

\author{Jun Zhang}
\affiliation{School of Physics and Technology, Wuhan University, Wuhan, Hubei 430072, China}
\affiliation{Beijing Computational Science Research Center, Beijing 100084, China}

\author{Yingying Han}
\affiliation{School of Physics and Technology, Wuhan University, Wuhan, Hubei 430072, China}

\author{Peng Xu}
\affiliation{School of Physics and Technology, Wuhan University, Wuhan, Hubei 430072, China}

\author{Wenxian Zhang}
\email[Corresponding email: ]{wxzhang@whu.edu.cn}
\affiliation{School of Physics and Technology, Wuhan University, Wuhan, Hubei 430072, China}

\date{\today}

\begin{abstract}
A challenge in precision measurement with squeezed spin state arises from the spin dephasing due to stray magnetic fields. To suppress such environmental noises, we employ a continuous driving protocol, rotary echo, to enhance the spin coherence of a spin-1 Bose-Einstein condensate in stray magnetic fields. Our analytical and numerical results show that the coherent and the squeezed spin states are preserved for a significantly long time, compared to the free induction decay time, if the condition $h\tau = m\pi$ is met with $h$ the pulse amplitude and $\tau$ pulse width. In particular, both the spin average and the spin squeezing, including the direction and the amplitude, are simultaneously fixed for a squeezed spin state. Our results point out a practical way to implement quantum measurements based on a spin-1 condensate beyond the standard quantum limit.

%From the analytical deviation, we also demonstrate that the suppression
%efficiency is independent of spin size.
\end{abstract}
%03.75.Mn	Multicomponent condensates; spinor condensates
%03.67.Pp	Quantum error correction and other methods for protection against decoherence
%76.60.Lz	Spin echoes
\maketitle

%%%%%%%%%%%%%%%%%%%%%%%%%%%%%%%%%%%%%%%%%%%%%%%%%%%%%%%%%%%%%%%%%%%%%%%%%%%%%%%
\section{Introduction}

Precision measurement with quantum entanglement and squeezed quantum state has the merit of measurement beyond the standard quantum limit~\cite{Giovannetti2011Advances,Muessel14PRL}. A spinor Bose-Einstein condensate (BEC) provides such a potential opportunity for quantum metrology with its wonderful controllability and fine-tuning in practical experiments~\cite{Ueda13RMP,Muessel14PRL,yixiaohuang12PRA,magnetometryBEC07,controlBEC03,PhysRevA.93.023616,Chang2005Coherent,Dunningham2012}. However, dephasing due to the stray magnetic fields in laboratories prevents the spinor BEC from wider applications and higher precisions, such as the quantum many-body ground state of an antiferromagnetic spin-1 BEC~\cite{Ueda13RMP,Chang04}. Suppressing the environmental noises such as the stray magnetic field becomes a long-standing challenge for quantum metrology and many-body quantum phase transition experiments utilizing spinor BECs~\cite{Eto13,eto2014control,Huang2015Spin,Dunningham2012,Chapman15,Marti2014}.

Many theoretical and experimental investigations have contributed to the suppressing of the stray magnetic fields. Some of these ideas are borrowed from the fields of nuclear magnetic resonance and quantum information and computation, such as the quantum error correction and dynamical decoupling with hard pulses~\cite{QEC2000,DDPRL1999,BBdymsupp}. For a spinor BEC system, there are also a few unique methods, such as the magnetic shield room, active compensation, and the spin Hahn echo pulse sequences~\cite{Ueda13RMP,Zhang10a,Hamley12,Hoang13,huanbinLi,zhang15,Eto13control,BYNing11}. These methods suppressed the stray magnetic fields to a much lower value, e.g., $1$ nT in a recent work by Eto \textit{et al}.~\cite{Eto13,eto2014control}. However, such a method is rather costly and the second order Zeeman effect of the strong driving field in the spin Hahn echo would prevent the method from perfect elimination of the noises~\cite{solomon1959rotary,2015Dobrovitski,continuousPRA,DobrovitskiPRB,zhang15}.

In this paper, we propose to use the dynamical decoupling protocols, rotary echo (RE), with continuous driving pulses in stead of the hard ones in $\delta$ shape (see Fig.~\ref{fig:echo} for a comparison), to avoid the error accumulation with time~\cite{solomon1959rotary,2015Dobrovitski,PhysRevB.92.060301, DobrovitskiPRB, continuousPRA,PhysRevB.84.161403,RE12,CDD2015PRA,Aiello2013Composite}.
Specially, for a spinor BEC, the main noise source includes the local phase noise from stray magnetic fields and the radio frequency intensity noise~\cite{klempt2011,klempt2015}. The contribution of the magnetic-field gradient is usually negligible due to the advanced experimental controllability and the smallness of the condensate ($\sim$10$\mu$m). We apply the RE protocol in a spin-1 BEC, which is dephased by stray magnetic fields, and evaluate the performance of the RE protocol for two interesting condensate spin states, the coherent spin state and the squeezed spin state. We find that the RE protocol extends significantly the coherent time of the system if the continuous-pulse duration is close to its magic values, $\tau = m\pi/h$ ($m=1,2,3,\cdots$) with $h$ the driving field strength. Our results point to a different direction to suppress the stray magnetic field effect in spinor BECs and to utilize the squeezed spin state in quantum metrology beyond the standard quantum limit~\cite{magmeter}.

The paper is organized as follows. In Sec.~\ref{sec:re}, we describe the spin-1 BEC system and the RE protocol, including the effect of the stray magnetic fields along $z$ axis and the driving field along $x$ axis. In Sec.~\ref{sec:rd}, we evaluate the performance of the RE protocol in preserving the spin average and the squeezing parameter for two interesting condensate spin states, the coherent spin state and the squeezed spin state. We give conclusions and discussions in Sec.~\ref{sec:con}.

%%%%%%%%%%%%%%%%%%%%%%%%%%%%%%%%%%%%%%%%%%%%%%%%%%%%%%%%%%%%%%%%%%%%%%%%%%%%%%%%%%%%%%%%%%
\section{System description and rotary echo protocol}
\label{sec:re}

\begin{figure}
\centering
\includegraphics[width=3in]{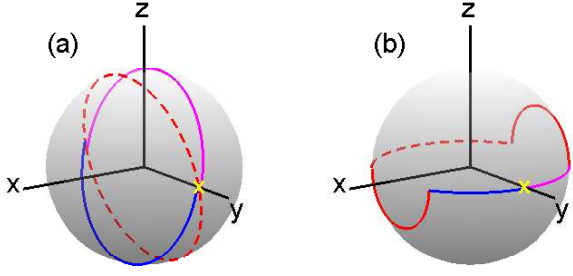}
\caption{\label{fig:echo} (Color online.) Typical spin trajectories in (a) rotary echo and (b) Hahn echo. The initial spin state is marked by a yellow cross. (a) In a rotary echo, the spin rotates about the total field $h\hat x + \omega \hat z$ during the first and the fourth quarter-cycle $\tau$, in purple and blue solid lines, respectively; the spin rotates about the total field $-h\hat x + \omega \hat z$ during the second and the third quarter-cycle, in red dashed lines. (b) In a Hahn echo, the spin evolves in a noise magnetic field $\omega \hat z$ for a cycle $4\tau$. At the moments $\tau$ and $3\tau$, the spin is rotated for a $\pi$ angle about $+x$ axis, denoted by red solid lines.}
\end{figure}

\begin{figure}
\centering
\includegraphics[width=3in]{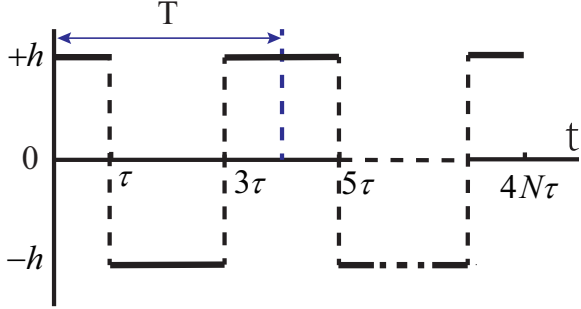}
\caption{\label{fig:rep} (Color online.) Pulse sequence of a RE. The RE is a symmetric sequence consisting of $N$ cycles. Within each cycle, the driving field direction is reversed between $+x$ and $-x$, denoted by $+h$ and $-h$, respectively. The two outer segments are along $+x$ for a duration time $\tau$ and the inner segment along $-x$ for a duration time $2\tau$. The period of a single cycle is $T=4\tau$. After $N$ cycles, the total time is $t=4N\tau$.}
\end{figure}

We consider a spin-1 condensate in a uniform magnetic field ${\bf B}$ with the Hamiltonian described by \cite{ho1998spinor,ohmi1998bose,Law98, Zhang03}
\begin{eqnarray}\label{eq:ha}
  \hat H &=& \int\mathrm{d}{\bf r}\left\{\hat\psi_{i}^{\dag}\left[-\frac{\hbar^{2}}{2M}\nabla^{2}+V_{ext}({\bf r})+E_Z\right]\hat\psi_{i} \right.\nonumber \\
   && \left.+ \frac{c_{0}}{2}\hat\psi_{i}^{\dag}\hat\psi_{j}^{\dag}\hat\psi_{j}\hat\psi_{i}
   +\frac{c_{2}}{2}\hat\psi_{k}^{\dag}\hat\psi_{i}^{\dag}(F_{\gamma})_{ij}(F_{\gamma})_{kl}\hat\psi_{j}\hat\psi_{l}\right\},
\end{eqnarray}
where $i,j,k,l \in \{\pm,0\}$ with $\pm,0$ denoting the magnetic quantum numbers $\pm 1,0$, respectively. Repeated indices are summed. $\hat \psi_{i}(\hat \psi_{i}^{\dag})$ is the field operator which annihilates (creates) an atom in the $i$th hyperfine state $|i\rangle \equiv |F=1,m_F=i\rangle$. $M$ is the mass of the atom. Interaction terms with coefficients $c_{0}$ and  $c_{2}$ describe elastic collisions of spin-1 atoms, namely, $c_{0}=4\pi \hbar^{2}(a_{0}+2a_{2})/3M$ and $c_{2}=4\pi \hbar^{2}(a_{2}-a_{0})/3M$ with $a_{0}$ and $a_{2}$ being the $s$-wave scattering lengths in singlet and quintuplet channels. The spin exchange interaction is ferromagnetic (anti-ferromagnetic) if $c_{2}<0 $ ($>0$). $F_{\gamma=x,y,z}$ are spin-1 matrices. $E_Z$ denotes the linear Zeeman shift of an alkali-metal atom,
$E_Z = -g_F \mu_B {\bf B}\cdot {\bf F}$,
where $g_{F}$ is the Land\'e $g$-factor for an atom with the total angular momentum $F=1$, such as $^{87}$Rb and $^{23}$Na atoms, and $\mu_{B}$ is the Bohr magneton.

We rewrite the Hamiltonian $\hat H$ as the sum of symmetric part $\hat H_s$ and the nonsymmetric part $\hat H_a$, i.e., $\hat H=\hat H_s+\hat H_a $, where
\begin{eqnarray}
\hat H_s&=&\int\mathrm{d}{\bf r}\hat\psi_{i}^{\dag}\left[-\frac{\hbar^{2}}{2M}\nabla^{2}+V({\bf r})\right]\hat\psi_{i}+ \frac{c_{0}}{2}\hat\psi_{i}^{\dag}\hat\psi_{j}^{\dag}\hat\psi_{j}\hat\psi_{i} \nonumber \\
\hat H_a&=&\int\mathrm{d}{\bf r}\left[\frac{c_{2}}{2}\hat\psi_{k}^{\dag}\hat\psi_{i}^{\dag}(F_{\gamma})_{ij}(F_{\gamma})_{kl}\hat\psi_{j}\hat\psi_{l} +\hat\psi_{i}^{\dag}E_Z\hat\psi_{i}\right]. \nonumber
\end{eqnarray}
For $^{87}$Rb or $^{23}$Na condensates, the density-dependent interaction $c_0$ is much larger than the spin-dependent interaction $c_2$; thus $\hat H_s$ is the dominate part. The condensate wave function $\phi_i({\bf r})$ are approximately described by the same wave function $\phi({\bf r})$, i.e., single-mode approximation (SMA). Such a wave function is defined by the Gross-Pitaevskii equation through $\hat H_s$,
\begin{equation}
\left(-\frac {\hbar^2}{2M}\nabla^2+V+c_0n \right)\phi({\bf r})=\mu\phi({\bf r}). \nonumber
\end{equation}
Under the condition of SMA, the field operator approximately becomes, at the zero temperature, $\hat\psi_i=\hat a_i\phi({\bf r})$, $i=0, \pm$. Note that $\hat a_i$ is the the annihilation operator associated with the condensate mode. So the leading part of $\hat H_s$ and $\hat H_a$ become~\cite{Law98}
\begin{widetext}
\begin{eqnarray}
\hat H_s&\approx&\mu \hat{N}-c_0'\hat{N}(\hat{N}-1)\equiv H_s \nonumber \\
\hat H_a&\approx& A(\hat{a}_1^{\dag}\hat{a}_1^{\dag}\hat{a}_1\hat{a}_1+\hat{a}_{-1}^{\dag}\hat{a}_{-1}^{\dag}\hat{a}_{-1}\hat{a}_{-1}-
2\hat{a}_1^{\dag}\hat{a}_{-1}^{\dag}\hat{a}_1\hat{a}_{-1}+2\hat{a}_1^{\dag}\hat{a}_0^{\dag}\hat{a}_1\hat{a}_0+
2\hat{a}_{-1}^{\dag}\hat{a}_0^{\dag}\hat{a}_{-1}\hat{a}_0+2\hat{a}_0^{\dag}\hat{a}_0^{\dag}\hat{a}_1\hat{a}_{-1}+
2\hat{a}_1^{\dag}\hat{a}_{-1}^{\dag}\hat{a}_0\hat{a}_0)\\
&& + \frac{\sqrt{2}}{2}b_x(\hat{a}_{0}^{\dag}\hat{a}_{1}+\hat{a}_{1}^{\dag}\hat{a}_{0}+\hat{a}_{-1}^{\dag}\hat{a}_{0}+\hat{a}_{0}^{\dag}\hat{a}_{-1})
+\frac{\sqrt{2}}{2i}b_y(\hat{a}_{1}^{\dag}\hat{a}_{0}+\hat{a}_{0}^{\dag}\hat{a}_{-1}-\hat{a}_{0}^{\dag}\hat{a}_{1}-\hat{a}_{-1}^{\dag}\hat{a}_{0})
+ b_z(\hat{a}_{-1}^{\dag}\hat{a}_{-1}-\hat{a}_{1}^{\dag}\hat{a}_{1})
\equiv H_a.\nonumber
\end{eqnarray}
\end{widetext}
Here $c_0'=(c_0/2)\int|\phi({\bf r})|^4\mathrm{d}{\bf r}$, $A=(c_2/2)\int|\phi({\bf r})|^4\mathrm{d}{\bf r}$, $\hat N=\hat{a}_1^{\dag}\hat{a}_1+\hat{a}_0^{\dag}\hat{a}_0+\hat{a}_{-1}^{\dag}\hat{a}_{-1}$
is the total number of the atoms, and $b_{x,y,z} = -g_F\mu_B B_{x,y,z}$.

By defining the angular momentum, i.e.,
$J_x=\sqrt{2}/2(\hat{a}_{0}^{\dag}\hat{a}_{1}+\hat{a}_{1}^{\dag}\hat{a}_{0}+\hat{a}_{-1}^{\dag}\hat{a}_{0}+\hat{a}_{0}^{\dag}\hat{a}_{-1})$,
$J_y=\sqrt{2}/({2i})(\hat{a}_{1}^{\dag}\hat{a}_{0}+\hat{a}_{0}^{\dag}\hat{a}_{-1}-\hat{a}_{0}^{\dag}\hat{a}_{1}-\hat{a}_{-1}^{\dag}\hat{a}_{0})$,
% $J_x=???\sqrt{2}(\hat{a}_{0}^{\dag}\hat{a}_{1}+\hat{a}_{-1}^{\dag}\hat{a}_{0})$,
%$J_y=???\sqrt{2}(\hat{a}_{1}^{\dag}\hat{a}_{0}+\hat{a}_{0}^{\dag}\hat{a}_{-1})$,
and $J_z=(\hat{a}_{-1}^{\dag}\hat{a}_{-1}-\hat{a}_{1}^{\dag}\hat{a}_{1})$, the Hamiltonian $H_a$ is rewritten in a simple form,
\begin{equation}~\label{eq:ha}
H_a=A({J^2}-2{N})+{\bf b}\cdot {\bf J} \nonumber
\end{equation}
with ${\bf b} = -g_F\mu_B {\bf B}$. The constant term $-2AN$ is omitted since the total atom number $N$ is conserved in experiments. Therefore, the Hamiltonian of the system becomes
\begin{equation}~\label{eq:eff}
H=AJ^2+{\bf b}\cdot {\bf J}.
\end{equation}
For a ferromagnetic spin-1 condensate, the SMA is valid in a wide range of magnetic fields~\cite{Yi02, Zhang03}. For such a ferromagnetic state, the total angular momentum $J$ equals to the total number of atoms $N$ at zero magnetic field. When the system is in a noisy environment, for example, in stray magnetic fields, the total magnetic field includes two sources: the applied external magnetic field and the stray magnetic field. In general, the external magnetic field is applied purposely and controlled accurately, but the stray magnetic field is noisy and needs to be suppressed since it usually causes the system to decohere in a short time.

In order to suppress the stray magnetic field and enhance the coherence of system, we employ the RE sequence as shown in Fig.~\ref{fig:echo}(a). Compared to the Hahn spin echo[Fig.~\ref{fig:echo}(b)], the RE shows significant advantages: besides the prevention of the error cumulation after many pulses, the applied magnetic field is much smaller than that in a Hahn echo and the quadratic Zeeman effect can be safely neglected.

For a typical RE protocol, the direction of the applied magnetic field (or the phase of driving) is periodically reversed between $+x$ and $-x$. As shown in Fig~\ref{fig:rep}, during each cycle, the driving field is along $+x$ for a duration of $\tau$, then turned to $-x$ for a  duration of $2\tau$, and turned back to $+x$ subsequently. The cycle period is $T=4\tau$. The complete Hamiltonian during a RE cycle is described by
\begin{equation}\label{eq:ham}
H_{\pm}=\pm hJ_{x}+AJ^{2}+ w J_{z},
\end{equation}
where $h$ is the amplitude of the driving field, $\pm$ indicate the direction of driving field. Here we have assumed the spin interaction is ferromagnetic $c_2<0$ and the minus sign before $H_{\pm}$ has been omitted without loss of generality. Hence, we consider $h=\mu_B g_F B_x, A$ and $w=\mu_B g_F B_z$ positive hereafter, where $B_x$ ($B_z$) is the strength of the driving (stray) magnetic field.

%%%%%%%%%%%%%%%%%%%%%%%%%%%%%%%%%%%%%%%%%%%%%%%%%%%%%%%%%%%%%%%%%%%%%%%%%%%%%%%%%%%%%%%%%%%%
\section{Suppressing the stray magnetic field effect with rotary echo}\label{sec:rd}

For a single cycle RE, the time evolution operator is
\begin{equation}\label{eq:evl}
U=e^{-i\tau H_{+}}e^{-2i\tau H_{-}}e^{-i\tau H_{+}}.
\end{equation}
For a given $w$, the evolution of the system is essentially an $SO(3)$ rotation in spin space; thus the evolution operator can be written as $\exp{(-i\theta J_n)}$, where $J_n=\bf J\cdot\bf n$ with $\bf n$ being the rotation axis and $\theta$ the rotation angle. From the Hamiltonian Eq. (\ref{eq:ham}), we find  $\theta=\tau\sqrt{h^2+w^2}$ and $J_{n,\pm}=\pm n_xJ_x+n_zJ_z$, where $n_x= h/\sqrt{h^2+w^2}$ and $n_z=w/\sqrt{h^2+w^2}$. Since it contributes nothing to the rotation, the term $AJ^2$ in the Hamiltonian Eq. (\ref{eq:ham}) is neglected here and reconsidered as needed. It is easy to find the rotation matrix of the outer and the inner segment of the RE cycle (shown in Fig.~\ref{fig:rep}), respectively,
\begin{eqnarray}
R_+&=&\left(
\begin{array}{ccc}
p_1 n_x^2+\cos \theta & -n_z \sin\theta & n_xn_z p_1  \\
 n_z\sin \theta & \cos \theta & -n_x \sin \theta\\
 n_xn_z p_1  & n_x \sin \theta & p_1 n_z^2+\cos \theta \\
\end{array}
\right),\nonumber\\
R_{-}&=&\left(
\begin{array}{ccc}
 p_2 n_x^2+\cos 2\theta & -n_z \sin 2\theta & -n_xn_z p_2 \\
 n_z\sin 2\theta & \cos 2\theta & n_x \sin 2\theta \\
 -n_xn_z p_2   & -n_x \sin 2\theta & p_2 n_z^2+\cos 2\theta \\
\end{array}
\right),\nonumber
\end{eqnarray}
where $p_1 = 1-\cos\theta$ and $p_2 = 1-\cos 2\theta$. As a RE cycle we consider has a time-reversal symmetry, which cancels all the odd terms proportional to $J_y$ in the average Hamiltonian~\cite{Haeberlen76}, only the even terms with $J_x$ and $J_z$ are left. Therefore, the total rotation within the RE cycle is $\tilde{U}=\exp(-i\alpha \tilde J_n)$ with $\tilde J_n=\tilde n_xJ_x+\tilde n_zJ_z$, and the rotation matrix becomes,
\begin{eqnarray}
\tilde R =\left(
\begin{array}{ccc}
q_1 \tilde n_x^2+\cos \alpha & -\tilde n_z \sin\alpha & \tilde n_x\tilde n_z q_1 \\
 \tilde n_z\sin \alpha & \cos \alpha & -\tilde n_x \sin \alpha \\
 \tilde n_x\tilde n_z q_1 & \tilde n_x \sin \alpha & q_1 \tilde n_z^2+\cos \alpha \\
\end{array}
\right)
\end{eqnarray}
with $q_1 = 1-\cos \alpha$. With the relationship $\tilde R=R_+R_-R_+$, the rotation angle $\alpha$ and the direction $\tilde n_x$ and $\tilde n_z$ are determined
\begin{eqnarray}
\cos \alpha &=& 1-n_z^4 (1-\cos 4 \theta)-4n_x^2n_z^2 (1-\cos 2 \theta)~\label{eq:alpha}\nonumber \\
\tilde n_x {\sin\alpha} &=& 8 n_xn_z^2 \sin ^2\left(\frac{\theta}{2}\right) \sin \theta \left(n_z^2 \cos 2\theta +n_x^2\right)\nonumber \\
\tilde n_z^2 &=& 1-\tilde n_x^2,
\end{eqnarray}
where we have defined $\cos\beta=n_x$ and $\sin\beta=n_z$.

The single cycle evolution operator Eq.(~\ref{eq:evl}) eventually becomes a single exponential
\begin{equation}~\label{eq:uni}
U=\exp[-i\alpha(\tilde n_xJ_x+\tilde n_zJ_z)] \exp(-i4\tau AJ^2).
\end{equation}
It is then straightforward to obtain the full $N$-cycle evolution operator
\begin{equation}~\label{eq:uni2}
U(t=4N\tau)=\exp[-i(\lambda J_x+\gamma J_z+4N\tau A J^2)],
\end{equation}
with $\lambda=N\alpha \tilde n_x$ and $\gamma=N\alpha \tilde n_z$.

In the analytical result Eq.~(\ref{eq:alpha}), if $\theta = m\pi$ ($m=1,2,3,\cdots$), $\cos\alpha=1$, thus $\alpha=0$ and the evolution operator $U$ becomes unity, which indicates that the given magnetic field effect is exactly canceled by the RE sequence.

Here we emphasize that the spin interaction term $AJ^2$ in Eq.~(\ref{eq:ham}) can not be excluded from the Hamiltonian of the BEC system, though it has no contribution to the results of our calculation. First, $AJ^2$ is a natural ingredient of the spinor BEC, which is often used to generate a special squeezed spin state, a twin-Fock state~\cite{klempt2011,klempt2015}. Second, we consider a ferromagnetic interaction $c_2<0$ so that a ferromagnetic coherent state (the ground state of the system) is stable and useful for a practical magnetometer. In addition, the RE driving field and the stray magnetic fields are small enough to omit their quadratic Zeeman effect; thus there is no necessity to consider the spin mixing dynamics of the driven spinor BEC~\cite{Zhang05a}.

In the following, we will numerically examine the performance of the RE sequence for two most interesting states: the coherent spin state (CSS) and the squeezed spin state (SSS). We will monitor the spin average for both states and the squeezing parameter for the SSS.

%%%%%%%%%%%%%%%%%%%%%%%%%%%%%%%%%%%%%%%%%%%%%%%%%%
\subsection{Coherent spin states}~\label{sec:A}

\begin{figure}
\centering
\includegraphics[width=3in]{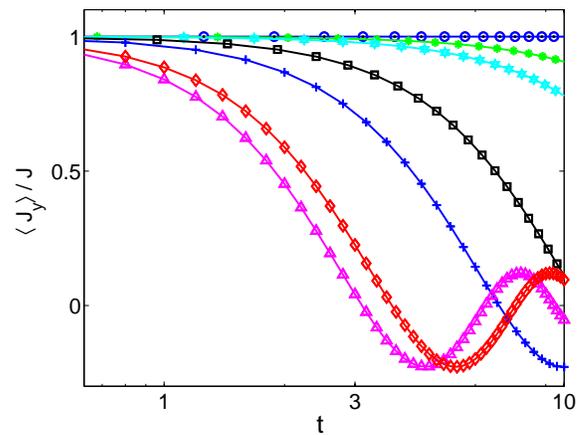}
\caption{\label{fig:expv}(Color online.) Spin average evolution under REs for $\tau=0.05$ (red diamonds), 0.1 (blue crosses), 0.12 (black squares), 0.14 (cyan hexagons), and 0.17 (green asterisks) with $h=20$. The blue circles are for $\tau = \pi/h$ with $h=20$. The FID is denoted with purple triangles. The solid lines, coinciding with the marks, are the corresponding analytical predictions by Eq.~(\ref{eq:fin}). The black dots(coinciding with the blue circles) are spin average under Hahn echoes as shown in Fig.~\ref{fig:echo}(b), where the cycle period is $4\tau = 0.2\pi$, the pulse amplitude is $10h=200$, and the pulse width is $0.005\pi$. Other parameters are $A=0.02$ and $J=100$. The average is performed over $N_r = 10^3$ random $w$'s distributed uniformly between $[0,1]$.
When $\tau=\pi/h$, the spin average barely decay under REs and Hahn echoes, indicating strong suppression of stray magnetic fields. }
\end{figure}

\begin{figure}
\centering
\includegraphics[width=3.3in]{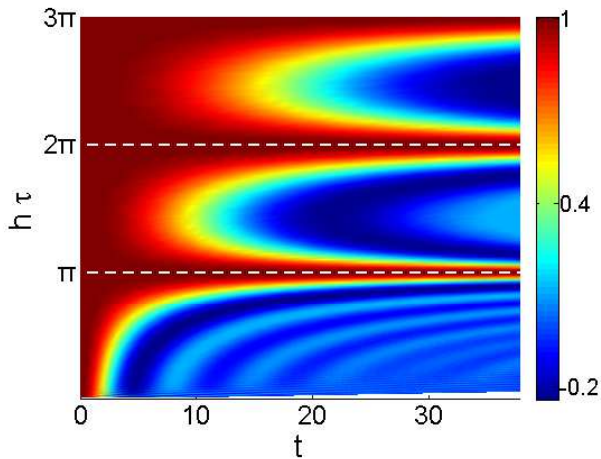}
\caption{\label{fig:ttau} (Color online) Dependence of the spin average on $t=4N\tau$ and $h\tau$. The parameters are $J=1$, $h=20$, and $A=0.02$. The spin average is conserved if $h\tau=m\pi \; (m=1, 2, 3, \cdots)$ as shown by horizontal dashed lines.}
\end{figure}

We consider an initial coherent spin state which is fully polarized along $y$ axis and evolves under the total Hamiltonian Eq.~(\ref{eq:ham}). We assume the stray magnetic-field effect as white noise, i.e., a series of random numbers ranging from 0 to $w_c$. We set the energy unit as $w_c=1$ and the corresponding time unit is $w_c^{-1}=1$. Other terms with $A$ and $h$ are all scaled with $w_c$. For a typical spin-1 BEC experiment, such as $^{87}$Rb, the stray magnetic fields are in the order of $1$ mG, i.e., roughly 3 kHz in energy. For a magnetic shield room with permalloy plates, the stray field can be further reduced down to $0.1$ mG~\cite{Eto13}.

The free evolution of the system, i.e., the free induction decay (FID), is often manifested by the decay of the transverse magnetization or the spin average
\begin{eqnarray}~\label{eq:sa}
\langle J_y\rangle (t) &=& \langle \psi(t) | J_y |\psi(t) \rangle
\end{eqnarray}
where $\psi(t)$ is the wave function of the condensate spin at time $t$. In the stray magnetic fields, the condensate spin only precesses around $z$ axis and is dephased with a characteristic time $T^*_2$. It is easy to find the FID signal
\begin{eqnarray}~\label{eq:sa}
\langle J_y\rangle (t) &=& J\;\frac {\sin(w_c t)} {w_c t} \nonumber
\end{eqnarray}
and the dephasing time is in the order of $T_2^* \sim w_c^{-1}$, as also shown in Fig.~\ref{fig:expv}.

By including the driving field $h$, the dephasing due to the stray magnetic field is suppressed by the REs, i.e., the spin average is larger than the FID signal. It is straightforward to calculate the spin average for a given $w$ at $t=4N\tau$
\begin{equation}
\langle J_y\rangle(t)=J\cos{(\sqrt{\lambda^2+\gamma^2}\,)}, \nonumber
\end{equation}
where we have adopted the $N$-cycle evolution operator in Eq.~(\ref{eq:uni2}). For a distribution of many $w$'s, we find
\begin{equation}~\label{eq:fin}
\langle J_y\rangle(t)=\frac J {N_r} \sum_{i=1}^{N_r} \cos{(\sqrt{\lambda_i^2+\gamma_i^2}\,)}
\end{equation}
with $N_r$ being the number of random samples. Since each stray magnetic field frequency $w_i$ is untrackable, we consider the strong driving regime, $h\gg w_c$, which is often the case in a practical precision measurement experiment. Hence, when $h\tau=m\pi$, $\langle J_y(t)\rangle$ barely decay.
%It clearly shows that $\langle J_y(t)\rangle$ barely decay when $h\tau=m\pi$ where $\beta_i$ and $\gamma_i$ both equal to zero. We note that the above analysis is valid for small $g\ll 1$.
To check the validity regime for $h\gg w_c$, we carry out numerical calculations and present both analytical and numerical results in Fig.~\ref{fig:expv}. For the parameters we choose ($g \lesssim 0.05$, where $g=w_c/h$), the analytical and numerical results agree well. As shown in Fig.~\ref{fig:expv}, the RE sequence always suppresses the stray magnetic-field effect and improves the spin average, no matter what value of $\tau$ is. As $\tau$ increases from a small value, the performance of the RE protocol becomes better, i.e., the spin average at a fixed time becomes larger. In particular, the spin average is almost a constant if the magic condition $\tau = \pi /h $ is satisfied [see also Fig.~\ref{fig:echo}(a)]. We remark that such a magic $\tau$ exists simultaneously for all $w_i$, instead of a single one. Once $\tau > \pi/h$, the spin average drops with $\tau$ increasing. In addition, our numerical results for $J=10$ and $J=100$ agree exactly, indicating that the performance of the RE protocol is independent of the spin size $J$. Finally, we compare the evolution of spin average under the magic REs and Hahn echoes with the same cycle period $4\tau = 0.2\pi$. The pulse amplitude and width of the Hahn echoes are, respectively, $10h=200$ and $0.1\tau=0.005\pi$. The performance of the magic RE protocol is essentially the same as that of the Hahn echo.

To investigate the effect of various $\tau$, we further calculate the spin average under RE sequences for different $\tau$s and present the results in Fig.~\ref{fig:ttau}. We see clear periodicity of $\tau = m\pi/h$ where the spin average is almost conserved, from Fig.~\ref{fig:ttau}. In addition, the good performance region, where the spin average is close to $J$ (red region in Fig.~\ref{fig:ttau}), becomes larger as $m$ increases.

%%%%%%%%%%%%%%%%%%%%%%%%%%%%%%%%%%%%%%%%%%%
\subsection{Squeezed spin states}~\label{sec:B}

\begin{figure}
\centering
\includegraphics[width=3in]{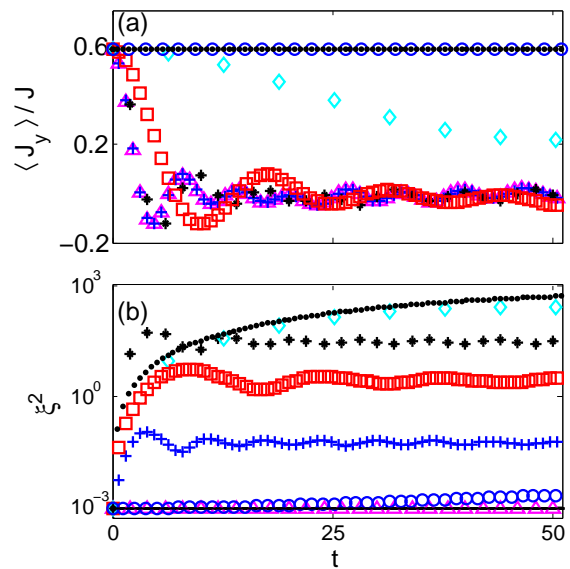}
\caption{\label{fig:sss} (Color online.) Time evolution of (a) the spin average and (b) the $z$-axis spin squeezing parameter for $J=10^3$. The initial direction of the spin is along $y$ axis and the initial optimal squeezing direction is along $z$ axis. The parameters are $h=2, \tau=0.1$ (blue crosses), $h=2, \tau=0.5$ (black asterisks), $h=2, \tau=\pi/h$ (cyan diamonds), $h=20, \tau=0.1$ (red squares),  and $h=20, \tau=\pi/h$ (blue circles). The FID is denoted with purple triangles. The black dots in (a) and (b) are the evolutions under the Hahn echoes with the cycle period of $4\tau = 0.2\pi$, the pulse amplitude of $10h = 200$, and the pulse width of $0.005\pi$. Other parameter is $A=0.02$. In (b), the black solid line (coinciding with the purple triangles) denotes the optimal squeezing parameter $\xi_s^2$ sought with numerical method. The spin average and the squeezing parameter are simultaneously fixed by the RE with $h\gg 1$ and $\tau = m\pi/h$. However, the Hahn echoes fail to fix the squeezing parameter.}
\end{figure}

We next consider the most interesting spin state, the SSS, which is widely used in precision measurement and quantum metrology. For such a spin state, besides its spin average $\langle J_y\rangle(t)$, we are also interested in the spin-squeezing parameter $\xi_{s}^2$, which represents the optimal or most squeezed transverse spin fluctuation~\cite{ma2011quantum, kitagawa1993squeezed,Chapman15},
\begin{equation}\label{eq:xis}
    \xi^2_s=\frac{2\min(\Delta J^2_{n_{\perp}})}{J},
\end{equation}
where $n_\perp$ denotes a transversal direction perpendicular to the spin average direction and the minimization (numerically) runs over all the transversal  directions. The squeezing parameter $\xi_s^2 = 1$ for a CSS and $\xi_s^2 < 1$ for a SSS.

A SSS can be generated by a one-axis or two-axis twisting Hamiltonian from an initial CSS with polarization along $y$ axis~\cite{RevModPhys.62.867,kitagawa1993squeezed,liu2011spin}. In this paper, we adopt the two-axis twisting method in which the optimal squeezing direction is invariant during the evolution~\cite{ma2011quantum}. The nonlinear two-axis twisting Hamiltonian is $H_{TAT}=\chi (J_xJ_z+J_zJ_x)$. The SSS is generated with spin average along $y$ axis, optimal squeezing direction along $z$ axis, and $\xi_s = 0.0009$ at $t = 0.002/\chi$ for $J = 10^3$.

After the preparation of the SSS, the $H_{TAT}$ is turned off. The spin state should be preserved without the stray magnetic fields. However, the SSS evolves by taking into account the stray magnetic fields. As shown in Fig.~\ref{fig:sss}, the spin average drops rapidly in the time scale of $w_c^{-1}$, but the spin squeezing parameter $\xi_s^2$ is not changed, nor is the optimal squeezing axis which is the same as the initial one.

By applying the driving field $h$ along $\pm x$ axis, both the spin average and the optimal squeezing axis are in general varying with time, but the optimal squeezing parameter $\xi_s^2$ is always unchanged, as shown in Fig.~\ref{fig:sss}. To better evaluate the performance of the RE protocol, we introduce the $z$-axis spin squeezing parameter
\begin{equation}\label{eq:xis}
    \xi^2=\frac{2\Delta J^2_z}{J}.
\end{equation}
For a perfect RE protocol, the initial SSS is preserved; thus both the spin average $\langle J_y\rangle$ and the $z$-axis spin squeezing parameter $\xi^2$ should be fixed.

We numerically simulate the time evolution of the condensate spin with the total Hamiltonian Eq. (\ref{eq:uni2}), starting from an initial SSS. The spin average, the $z$-axis spin squeezing parameter $\xi^2$, and the optimal spin squeezing parameter $\xi_s^2$ are presented in Fig.~\ref{fig:sss}. The weak driving $h=2$ ($g\lesssim 0.5$) case and the strong driving $h=20$ ($g\lesssim 0.05$) case are both included.

In the cases of weak driving, the spin average increases with $\tau$ increasing at a fixed time, indicating that the RE protocol works for preserving the spin average. However, the spin average still drops significantly in an intermediate time even at the magic condition $\tau = \pi/h$. %possibly due to the high order terms ${\cal O}(g^3)$.
Furthermore, the squeezing parameter $\xi^2$ increases to pretty large values for all the weak driving cases. The larger the $\tau$ is, the bigger the squeezing parameter $\xi^2$ is. Therefore, the weak driving RE protocol is not good at preserving the SSS and suppressing the stray magnetic-field effect.

In the cases of strong driving, the spin average at a fixed time becomes larger with a bigger $\tau$. Particularly, the spin average is almost a constant at the magic condition $\tau = \pi/h$. More remarkably, the squeezing parameter $\xi^2$ increases only slightly at times much longer than the dephasing time $T_2^* \sim w_c^{-1}$. Our numerical results confirm that the RE protocol with strong driving preserves not only the spin average but also the spin squeezing, including the amplitude and the direction. However, as a comparison, the Hahn echo protocol only preserves the spin average rather than the spin squeezing parameter. Therefore, the stray magnetic-field effect can be significantly suppressed by the RE protocol at magic condition $\tau = m\pi /h$ with $h\gg w_c$.

%%%%%%%%%%%%%%%%%%%%%%%%%%%%%%%%%%%%%%%%%%%%%%%%%%%%%%%%%%%%%%%%%%%%%%%%%%%%%%%%%%%%%%%%%%%
\section{Conclusion}\label{sec:con}

We investigate with analytical and numerical methods the performance of the rotary echo protocol in suppressing the effects of the stray magnetic fields. Our results show that the RE protocol with strong driving field $h$ at magic condition $\tau = m\pi /h$ well preserves the spin average for a coherent spin state of a spin-1 Bose condensate. Moreover, for a squeezed spin state, the same magic RE protocol simultaneously preserves not only the spin average but also the spin squeezing, including the amplitude and direction. Such a consequence of the RE protocol on spin squeezing has not been well appreciated in previous investigations in nuclear magnetic resonance community. The effects of the stray magnetic fields are significantly suppressed by the magic RE protocol. To implement the RE protocol in spin-1 BEC experiments, the applied driving field can be as small as $\sim 1$ mG for a magnetic shield room and $\sim 10$ mG for an ordinary laboratory, much smaller than the field applied in standard Hahn spin echo experiments ($\sim 100$ mG)~\cite{Eto13,eto2014control,zhang15, Ueda13RMP}. Our results provide a practical method to control the
stray magnetic-field effect in spinor BEC experiments and may find wide applications in precision measurement and quantum metrology, quantum phase transition, and ground-state properties in spinor Bose condensates~\cite{magmeter, Ueda13RMP, Pu16}.

\begin{acknowledgments}
W.Z. thanks V. V. Dobrovitski and M. S. Chapman for inspiring discussions. W.Z. is grateful to Beijing CSRC for hospitality. This work is supported by the National Natural Science Foundation of China under Grants No. 11574239, No.11275139, and No.11547310, the National Basic Research Program of China (Grant No. 2013CB922003), and the Fundamental Research Funds for the Central Universities.
\end{acknowledgments}

\appendix

\section{ RE protocol with colored noises}
To check the performance of the RE protocol under different types of noise, we evaluate the spin average and spin squeezing for a SSS in the presence of $1/f$ noise and Gaussian noise, besides the previous white noise. The $1/f$ noise is numerically distributed in the range $[4\times10^{-5}, 1]w_c$, while the Gaussian noise is in $[0,w_c]$ with an average of $0.5w_c$ and a full width at half maximum $0.1w_c$. The distributions of the noises are normalized and contain $1000$ samples in our calculation. Other parameters are the same as Fig.~\ref{fig:sss}.

\begin{figure}[H]
%\centering
\includegraphics[width=3in]{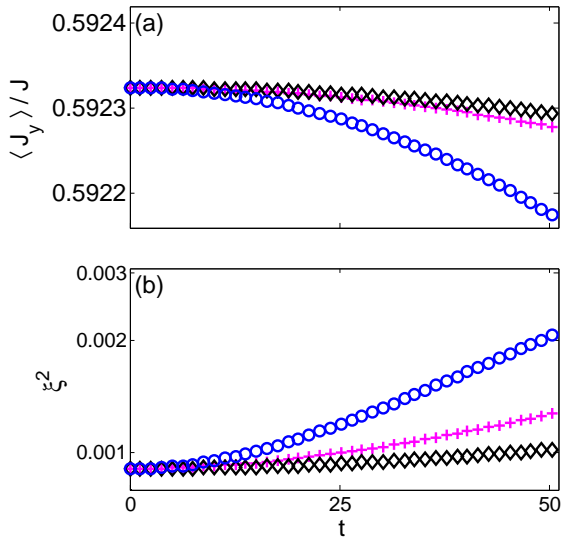}% Here is how to import EPS art
\caption{\label{fig:cn} (Color online.) Time evolution of (a) the spin average
and (b) the z-axis spin squeezing parameter in the presence of $1/f$ noise (purple crosses), Gaussian noise (black diamonds), and white noise (blue circles). Other parameters are the same as in Fig.~\ref{fig:sss}. The suppression of the $1/f$ and the Gaussian noise is more efficient than that of the white noise.}
\end{figure}

The comparisons are shown in Fig.~\ref{fig:cn}. Clearly, the RE protocol also suppresses the colored noises and the performances of the RE protocol in the colored noises are actually better than in the white noise, i.e., the spin averages are higher and the spin squeezing parameters are lower.


\begin{thebibliography}{49}
\expandafter\ifx\csname natexlab\endcsname\relax\def\natexlab#1{#1}\fi
\expandafter\ifx\csname bibnamefont\endcsname\relax
  \def\bibnamefont#1{#1}\fi
\expandafter\ifx\csname bibfnamefont\endcsname\relax
  \def\bibfnamefont#1{#1}\fi
\expandafter\ifx\csname citenamefont\endcsname\relax
  \def\citenamefont#1{#1}\fi
\expandafter\ifx\csname url\endcsname\relax
  \def\url#1{\texttt{#1}}\fi
\expandafter\ifx\csname urlprefix\endcsname\relax\def\urlprefix{URL }\fi
\providecommand{\bibinfo}[2]{#2}
\providecommand{\eprint}[2][]{\url{#2}}

\bibitem[{\citenamefont{Giovannetti et~al.}(2011)\citenamefont{Giovannetti,
  Lloyd, and Maccone}}]{Giovannetti2011Advances}
\bibinfo{author}{\bibfnamefont{V.}~\bibnamefont{Giovannetti}},
  \bibinfo{author}{\bibfnamefont{S.}~\bibnamefont{Lloyd}}, \bibnamefont{and}
  \bibinfo{author}{\bibfnamefont{L.}~\bibnamefont{Maccone}},
  \bibinfo{journal}{Nature Photon.} \textbf{\bibinfo{volume}{5}},
  \bibinfo{pages}{222} (\bibinfo{year}{2011}).

\bibitem[{\citenamefont{Muessel et~al.}(2014)\citenamefont{Muessel, Strobel,
  Linnemann, Hume, and Oberthaler}}]{Muessel14PRL}
\bibinfo{author}{\bibfnamefont{W.}~\bibnamefont{Muessel}},
  \bibinfo{author}{\bibfnamefont{H.}~\bibnamefont{Strobel}},
  \bibinfo{author}{\bibfnamefont{D.}~\bibnamefont{Linnemann}},
  \bibinfo{author}{\bibfnamefont{D.~B.} \bibnamefont{Hume}}, \bibnamefont{and}
  \bibinfo{author}{\bibfnamefont{M.~K.} \bibnamefont{Oberthaler}},
  \bibinfo{journal}{Phys. Rev. Lett.} \textbf{\bibinfo{volume}{113}},
  \bibinfo{pages}{103004} (\bibinfo{year}{2014}).

\bibitem[{\citenamefont{Stamper-Kurn and Ueda}(2013)}]{Ueda13RMP}
\bibinfo{author}{\bibfnamefont{D.~M.} \bibnamefont{Stamper-Kurn}}
  \bibnamefont{and} \bibinfo{author}{\bibfnamefont{M.}~\bibnamefont{Ueda}},
  \bibinfo{journal}{Rev. Mod. Phys.} \textbf{\bibinfo{volume}{85}},
  \bibinfo{pages}{1191} (\bibinfo{year}{2013}).

\bibitem[{\citenamefont{Huang et~al.}(2012)\citenamefont{Huang, Zhang, L\"u,
  Wang, and Yi}}]{yixiaohuang12PRA}
\bibinfo{author}{\bibfnamefont{Y.}~\bibnamefont{Huang}},
  \bibinfo{author}{\bibfnamefont{Y.}~\bibnamefont{Zhang}},
  \bibinfo{author}{\bibfnamefont{R.}~\bibnamefont{L\"u}},
  \bibinfo{author}{\bibfnamefont{X.}~\bibnamefont{Wang}}, \bibnamefont{and}
  \bibinfo{author}{\bibfnamefont{S.}~\bibnamefont{Yi}}, \bibinfo{journal}{Phys.
  Rev. A} \textbf{\bibinfo{volume}{86}}, \bibinfo{pages}{043625}
  (\bibinfo{year}{2012}).

\bibitem[{\citenamefont{Vengalattore et~al.}(2007)\citenamefont{Vengalattore,
  Higbie, Leslie, Guzman, Sadler, and Stamper-Kurn}}]{magnetometryBEC07}
\bibinfo{author}{\bibfnamefont{M.}~\bibnamefont{Vengalattore}},
  \bibinfo{author}{\bibfnamefont{J.~M.} \bibnamefont{Higbie}},
  \bibinfo{author}{\bibfnamefont{S.~R.} \bibnamefont{Leslie}},
  \bibinfo{author}{\bibfnamefont{J.}~\bibnamefont{Guzman}},
  \bibinfo{author}{\bibfnamefont{L.~E.} \bibnamefont{Sadler}},
  \bibnamefont{and} \bibinfo{author}{\bibfnamefont{D.~M.}
  \bibnamefont{Stamper-Kurn}}, \bibinfo{journal}{Phys. Rev. Lett.}
  \textbf{\bibinfo{volume}{98}}, \bibinfo{pages}{200801}
  (\bibinfo{year}{2007}).

\bibitem[{\citenamefont{Abdullaev et~al.}(2003)\citenamefont{Abdullaev, Caputo,
  Kraenkel, and Malomed}}]{controlBEC03}
\bibinfo{author}{\bibfnamefont{F.~K.} \bibnamefont{Abdullaev}},
  \bibinfo{author}{\bibfnamefont{J.~G.} \bibnamefont{Caputo}},
  \bibinfo{author}{\bibfnamefont{R.~A.} \bibnamefont{Kraenkel}},
  \bibnamefont{and} \bibinfo{author}{\bibfnamefont{B.~A.}
  \bibnamefont{Malomed}}, \bibinfo{journal}{Phys. Rev. A}
  \textbf{\bibinfo{volume}{67}}, \bibinfo{pages}{013605}
  (\bibinfo{year}{2003}).

\bibitem[{\citenamefont{Nolan et~al.}(2016)\citenamefont{Nolan, Sabbatini,
  Bromley, Davis, and Haine}}]{PhysRevA.93.023616}
\bibinfo{author}{\bibfnamefont{S.~P.} \bibnamefont{Nolan}},
  \bibinfo{author}{\bibfnamefont{J.}~\bibnamefont{Sabbatini}},
  \bibinfo{author}{\bibfnamefont{M.~W.~J.} \bibnamefont{Bromley}},
  \bibinfo{author}{\bibfnamefont{M.~J.} \bibnamefont{Davis}}, \bibnamefont{and}
  \bibinfo{author}{\bibfnamefont{S.~A.} \bibnamefont{Haine}},
  \bibinfo{journal}{Phys. Rev. A} \textbf{\bibinfo{volume}{93}},
  \bibinfo{pages}{023616} (\bibinfo{year}{2016}).

\bibitem[{\citenamefont{Chang et~al.}(2005)\citenamefont{Chang, Qin, Zhang,
  You, and Chapman}}]{Chang2005Coherent}
\bibinfo{author}{\bibfnamefont{M.~S.} \bibnamefont{Chang}},
  \bibinfo{author}{\bibfnamefont{Q.}~\bibnamefont{Qin}},
  \bibinfo{author}{\bibfnamefont{W.}~\bibnamefont{Zhang}},
  \bibinfo{author}{\bibfnamefont{L.}~\bibnamefont{You}}, \bibnamefont{and}
  \bibinfo{author}{\bibfnamefont{M.~S.} \bibnamefont{Chapman}},
  \bibinfo{journal}{Nature Phys.} \textbf{\bibinfo{volume}{1}},
  \bibinfo{pages}{111} (\bibinfo{year}{2005}).

\bibitem[{\citenamefont{Dunningham et~al.}(2012)\citenamefont{Dunningham,
  Cooper, and Hallwood}}]{Dunningham2012}
\bibinfo{author}{\bibfnamefont{J.~A.} \bibnamefont{Dunningham}},
  \bibinfo{author}{\bibfnamefont{J.~J.} \bibnamefont{Cooper}},
  \bibnamefont{and} \bibinfo{author}{\bibfnamefont{D.~W.}
  \bibnamefont{Hallwood}}, \bibinfo{journal}{AIP Conference Proceedings}
  \textbf{\bibinfo{volume}{1469}} (\bibinfo{year}{2012}).

\bibitem[{\citenamefont{Chang et~al.}(2004)\citenamefont{Chang, Hamley,
  Barrett, Sauer, Fortier, Zhang, You, and Chapman}}]{Chang04}
\bibinfo{author}{\bibfnamefont{M.-S.} \bibnamefont{Chang}},
  \bibinfo{author}{\bibfnamefont{C.~D.} \bibnamefont{Hamley}},
  \bibinfo{author}{\bibfnamefont{M.~D.} \bibnamefont{Barrett}},
  \bibinfo{author}{\bibfnamefont{J.~A.} \bibnamefont{Sauer}},
  \bibinfo{author}{\bibfnamefont{K.~M.} \bibnamefont{Fortier}},
  \bibinfo{author}{\bibfnamefont{W.}~\bibnamefont{Zhang}},
  \bibinfo{author}{\bibfnamefont{L.}~\bibnamefont{You}}, \bibnamefont{and}
  \bibinfo{author}{\bibfnamefont{M.~S.} \bibnamefont{Chapman}},
  \bibinfo{journal}{Phys. Rev. Lett.} \textbf{\bibinfo{volume}{92}},
  \bibinfo{pages}{140403} (\bibinfo{year}{2004}).

\bibitem[{\citenamefont{Eto et~al.}(2013{\natexlab{a}})\citenamefont{Eto,
  Ikeda, Suzuki, Hasegawa, Tomiyama, Sekine, Sadgrove, and Hirano}}]{Eto13}
\bibinfo{author}{\bibfnamefont{Y.}~\bibnamefont{Eto}},
  \bibinfo{author}{\bibfnamefont{H.}~\bibnamefont{Ikeda}},
  \bibinfo{author}{\bibfnamefont{H.}~\bibnamefont{Suzuki}},
  \bibinfo{author}{\bibfnamefont{S.}~\bibnamefont{Hasegawa}},
  \bibinfo{author}{\bibfnamefont{Y.}~\bibnamefont{Tomiyama}},
  \bibinfo{author}{\bibfnamefont{S.}~\bibnamefont{Sekine}},
  \bibinfo{author}{\bibfnamefont{M.}~\bibnamefont{Sadgrove}}, \bibnamefont{and}
  \bibinfo{author}{\bibfnamefont{T.}~\bibnamefont{Hirano}},
  \bibinfo{journal}{Phys. Rev. A} \textbf{\bibinfo{volume}{88}},
  \bibinfo{pages}{031602} (\bibinfo{year}{2013}{\natexlab{a}}).

\bibitem[{\citenamefont{Eto et~al.}(2014)\citenamefont{Eto, Sadgrove, Hasegawa,
  Saito, and Hirano}}]{eto2014control}
\bibinfo{author}{\bibfnamefont{Y.}~\bibnamefont{Eto}},
  \bibinfo{author}{\bibfnamefont{M.}~\bibnamefont{Sadgrove}},
  \bibinfo{author}{\bibfnamefont{S.}~\bibnamefont{Hasegawa}},
  \bibinfo{author}{\bibfnamefont{H.}~\bibnamefont{Saito}}, \bibnamefont{and}
  \bibinfo{author}{\bibfnamefont{T.}~\bibnamefont{Hirano}},
  \bibinfo{journal}{Phys. Rev. A} \textbf{\bibinfo{volume}{90}},
  \bibinfo{pages}{013626} (\bibinfo{year}{2014}).

\bibitem[{\citenamefont{Huang and Hu}(2015)}]{Huang2015Spin}
\bibinfo{author}{\bibfnamefont{Y.}~\bibnamefont{Huang}} \bibnamefont{and}
  \bibinfo{author}{\bibfnamefont{Z.~D.} \bibnamefont{Hu}},
  \bibinfo{journal}{Sci. Rep.} \textbf{\bibinfo{volume}{5}},
  \bibinfo{pages}{8006} (\bibinfo{year}{2015}).

\bibitem[{Cha()}]{Chapman15}
\bibinfo{note}{T. M.,Hoang and M.,Anquez and B. A., Robbins and X. Y., Yang and
  B. J., Land and C. D., Hamley and M. S., Chapman, arXiv:1512.05645}.

\bibitem[{\citenamefont{Marti et~al.}(2014)\citenamefont{Marti, MacRae, Olf,
  Lourette, Fang, and Stamper-Kurn}}]{Marti2014}
\bibinfo{author}{\bibfnamefont{G.~E.} \bibnamefont{Marti}},
  \bibinfo{author}{\bibfnamefont{A.}~\bibnamefont{MacRae}},
  \bibinfo{author}{\bibfnamefont{R.}~\bibnamefont{Olf}},
  \bibinfo{author}{\bibfnamefont{S.}~\bibnamefont{Lourette}},
  \bibinfo{author}{\bibfnamefont{F.}~\bibnamefont{Fang}}, \bibnamefont{and}
  \bibinfo{author}{\bibfnamefont{D.~M.} \bibnamefont{Stamper-Kurn}},
  \bibinfo{journal}{Phys. Rev. Lett.} \textbf{\bibinfo{volume}{113}},
  \bibinfo{pages}{155302} (\bibinfo{year}{2014}).

\bibitem[{\citenamefont{Knill et~al.}(2000)\citenamefont{Knill, Laflamme, and
  Viola}}]{QEC2000}
\bibinfo{author}{\bibfnamefont{E.}~\bibnamefont{Knill}},
  \bibinfo{author}{\bibfnamefont{R.}~\bibnamefont{Laflamme}}, \bibnamefont{and}
  \bibinfo{author}{\bibfnamefont{L.}~\bibnamefont{Viola}},
  \bibinfo{journal}{Phys. Rev. Lett.} \textbf{\bibinfo{volume}{84}},
  \bibinfo{pages}{2525} (\bibinfo{year}{2000}).

\bibitem[{\citenamefont{Viola et~al.}(1999)\citenamefont{Viola, Knill, and
  Lloyd}}]{DDPRL1999}
\bibinfo{author}{\bibfnamefont{L.}~\bibnamefont{Viola}},
  \bibinfo{author}{\bibfnamefont{E.}~\bibnamefont{Knill}}, \bibnamefont{and}
  \bibinfo{author}{\bibfnamefont{S.}~\bibnamefont{Lloyd}},
  \bibinfo{journal}{Phys. Rev. Lett.} \textbf{\bibinfo{volume}{82}},
  \bibinfo{pages}{2417} (\bibinfo{year}{1999}).

\bibitem[{\citenamefont{Viola and Lloyd}(1998)}]{BBdymsupp}
\bibinfo{author}{\bibfnamefont{L.}~\bibnamefont{Viola}} \bibnamefont{and}
  \bibinfo{author}{\bibfnamefont{S.}~\bibnamefont{Lloyd}},
  \bibinfo{journal}{Phys. Rev. A} \textbf{\bibinfo{volume}{58}},
  \bibinfo{pages}{2733} (\bibinfo{year}{1998}).

\bibitem[{\citenamefont{Zhang et~al.}(2010)\citenamefont{Zhang, Sun, Chapman,
  and You}}]{Zhang10a}
\bibinfo{author}{\bibfnamefont{W.}~\bibnamefont{Zhang}},
  \bibinfo{author}{\bibfnamefont{B.}~\bibnamefont{Sun}},
  \bibinfo{author}{\bibfnamefont{M.~S.} \bibnamefont{Chapman}},
  \bibnamefont{and} \bibinfo{author}{\bibfnamefont{L.}~\bibnamefont{You}},
  \bibinfo{journal}{Phys. Rev. A} \textbf{\bibinfo{volume}{81}},
  \bibinfo{pages}{033602} (\bibinfo{year}{2010}).

\bibitem[{\citenamefont{Hamley et~al.}(2012)\citenamefont{Hamley, Gerving,
  Hoang, Bookjans, and Chapman}}]{Hamley12}
\bibinfo{author}{\bibfnamefont{C.~D.} \bibnamefont{Hamley}},
  \bibinfo{author}{\bibfnamefont{C.~S.} \bibnamefont{Gerving}},
  \bibinfo{author}{\bibfnamefont{T.~M.} \bibnamefont{Hoang}},
  \bibinfo{author}{\bibfnamefont{E.~M.} \bibnamefont{Bookjans}},
  \bibnamefont{and} \bibinfo{author}{\bibfnamefont{M.~S.}
  \bibnamefont{Chapman}}, \bibinfo{journal}{Nature Phys.}
  \textbf{\bibinfo{volume}{8}}, \bibinfo{pages}{305} (\bibinfo{year}{2012}).

\bibitem[{\citenamefont{Hoang et~al.}(2013)\citenamefont{Hoang, Gerving, Land,
  Anquez, Hamley, and Chapman}}]{Hoang13}
\bibinfo{author}{\bibfnamefont{T.~M.} \bibnamefont{Hoang}},
  \bibinfo{author}{\bibfnamefont{C.~S.} \bibnamefont{Gerving}},
  \bibinfo{author}{\bibfnamefont{B.~J.} \bibnamefont{Land}},
  \bibinfo{author}{\bibfnamefont{M.}~\bibnamefont{Anquez}},
  \bibinfo{author}{\bibfnamefont{C.~D.} \bibnamefont{Hamley}},
  \bibnamefont{and} \bibinfo{author}{\bibfnamefont{M.~S.}
  \bibnamefont{Chapman}}, \bibinfo{journal}{Phys. Rev. Lett.}
  \textbf{\bibinfo{volume}{111}}, \bibinfo{pages}{090403}
  (\bibinfo{year}{2013}).

\bibitem[{\citenamefont{Li et~al.}(2015)\citenamefont{Li, Pu, Chapman, and
  Zhang}}]{huanbinLi}
\bibinfo{author}{\bibfnamefont{H.}~\bibnamefont{Li}},
  \bibinfo{author}{\bibfnamefont{Z.}~\bibnamefont{Pu}},
  \bibinfo{author}{\bibfnamefont{M.~S.} \bibnamefont{Chapman}},
  \bibnamefont{and} \bibinfo{author}{\bibfnamefont{W.}~\bibnamefont{Zhang}},
  \bibinfo{journal}{Phys. Rev. A} \textbf{\bibinfo{volume}{92}},
  \bibinfo{pages}{013630} (\bibinfo{year}{2015}).

\bibitem[{\citenamefont{Zhang et~al.}(2015)\citenamefont{Zhang, Yi, Chapman,
  and You}}]{zhang15}
\bibinfo{author}{\bibfnamefont{W.}~\bibnamefont{Zhang}},
  \bibinfo{author}{\bibfnamefont{S.}~\bibnamefont{Yi}},
  \bibinfo{author}{\bibfnamefont{M.~S.} \bibnamefont{Chapman}},
  \bibnamefont{and} \bibinfo{author}{\bibfnamefont{J.~Q.} \bibnamefont{You}},
  \bibinfo{journal}{Phys. Rev. A} \textbf{\bibinfo{volume}{92}},
  \bibinfo{pages}{023615} (\bibinfo{year}{2015}).

\bibitem[{\citenamefont{Eto et~al.}(2013{\natexlab{b}})\citenamefont{Eto,
  Sekine, Hasegawa, Sadgrove, Saito, and Hirano}}]{Eto13control}
\bibinfo{author}{\bibfnamefont{Y.}~\bibnamefont{Eto}},
  \bibinfo{author}{\bibfnamefont{S.}~\bibnamefont{Sekine}},
  \bibinfo{author}{\bibfnamefont{S.}~\bibnamefont{Hasegawa}},
  \bibinfo{author}{\bibfnamefont{M.}~\bibnamefont{Sadgrove}},
  \bibinfo{author}{\bibfnamefont{H.}~\bibnamefont{Saito}}, \bibnamefont{and}
  \bibinfo{author}{\bibfnamefont{T.}~\bibnamefont{Hirano}},
  \bibinfo{journal}{Appl. Phys. Express} \textbf{\bibinfo{volume}{6}},
  \bibinfo{pages}{052801} (\bibinfo{year}{2013}{\natexlab{b}}).

\bibitem[{\citenamefont{Ning et~al.}(2011)\citenamefont{Ning, Zhuang, You, and
  Zhang}}]{BYNing11}
\bibinfo{author}{\bibfnamefont{B.-Y.} \bibnamefont{Ning}},
  \bibinfo{author}{\bibfnamefont{J.}~\bibnamefont{Zhuang}},
  \bibinfo{author}{\bibfnamefont{J.~Q.} \bibnamefont{You}}, \bibnamefont{and}
  \bibinfo{author}{\bibfnamefont{W.}~\bibnamefont{Zhang}},
  \bibinfo{journal}{Phys. Rev. A} \textbf{\bibinfo{volume}{84}},
  \bibinfo{pages}{013606} (\bibinfo{year}{2011}).

\bibitem[{\citenamefont{Solomon}(1959)}]{solomon1959rotary}
\bibinfo{author}{\bibfnamefont{I.}~\bibnamefont{Solomon}},
  \bibinfo{journal}{Phys. Rev. Lett.} \textbf{\bibinfo{volume}{2}},
  \bibinfo{pages}{301} (\bibinfo{year}{1959}).

\bibitem[{\citenamefont{Mkhitaryan et~al.}(2015)\citenamefont{Mkhitaryan,
  Jelezko, and Dobrovitski}}]{2015Dobrovitski}
\bibinfo{author}{\bibfnamefont{V.}~\bibnamefont{Mkhitaryan}},
  \bibinfo{author}{\bibfnamefont{F.}~\bibnamefont{Jelezko}}, \bibnamefont{and}
  \bibinfo{author}{\bibfnamefont{V.}~\bibnamefont{Dobrovitski}},
  \bibinfo{journal}{Sci. Rep.} \textbf{\bibinfo{volume}{5}},
  \bibinfo{pages}{15402} (\bibinfo{year}{2015}).

\bibitem[{\citenamefont{Hirose et~al.}(2012)\citenamefont{Hirose, Aiello, and
  Cappellaro}}]{continuousPRA}
\bibinfo{author}{\bibfnamefont{M.}~\bibnamefont{Hirose}},
  \bibinfo{author}{\bibfnamefont{C.~D.} \bibnamefont{Aiello}},
  \bibnamefont{and}
  \bibinfo{author}{\bibfnamefont{P.}~\bibnamefont{Cappellaro}},
  \bibinfo{journal}{Phys. Rev. A} \textbf{\bibinfo{volume}{86}},
  \bibinfo{pages}{062320} (\bibinfo{year}{2012}).

\bibitem[{\citenamefont{Mkhitaryan and Dobrovitski}(2014)}]{DobrovitskiPRB}
\bibinfo{author}{\bibfnamefont{V.~V.} \bibnamefont{Mkhitaryan}}
  \bibnamefont{and} \bibinfo{author}{\bibfnamefont{V.~V.}
  \bibnamefont{Dobrovitski}}, \bibinfo{journal}{Phys. Rev. B}
  \textbf{\bibinfo{volume}{89}}, \bibinfo{pages}{224402}
  (\bibinfo{year}{2014}).

\bibitem[{\citenamefont{Farfurnik et~al.}(2015)\citenamefont{Farfurnik,
  Jarmola, Pham, Wang, Dobrovitski, Walsworth, Budker, and
  Bar-Gill}}]{PhysRevB.92.060301}
\bibinfo{author}{\bibfnamefont{D.}~\bibnamefont{Farfurnik}},
  \bibinfo{author}{\bibfnamefont{A.}~\bibnamefont{Jarmola}},
  \bibinfo{author}{\bibfnamefont{L.~M.} \bibnamefont{Pham}},
  \bibinfo{author}{\bibfnamefont{Z.~H.} \bibnamefont{Wang}},
  \bibinfo{author}{\bibfnamefont{V.~V.} \bibnamefont{Dobrovitski}},
  \bibinfo{author}{\bibfnamefont{R.~L.} \bibnamefont{Walsworth}},
  \bibinfo{author}{\bibfnamefont{D.}~\bibnamefont{Budker}}, \bibnamefont{and}
  \bibinfo{author}{\bibfnamefont{N.}~\bibnamefont{Bar-Gill}},
  \bibinfo{journal}{Phys. Rev. B} \textbf{\bibinfo{volume}{92}},
  \bibinfo{pages}{060301} (\bibinfo{year}{2015}).

\bibitem[{\citenamefont{Laraoui and Meriles}(2011)}]{PhysRevB.84.161403}
\bibinfo{author}{\bibfnamefont{A.}~\bibnamefont{Laraoui}} \bibnamefont{and}
  \bibinfo{author}{\bibfnamefont{C.~A.} \bibnamefont{Meriles}},
  \bibinfo{journal}{Phys. Rev. B} \textbf{\bibinfo{volume}{84}},
  \bibinfo{pages}{161403} (\bibinfo{year}{2011}).

\bibitem[{\citenamefont{Gustavsson et~al.}(2012)\citenamefont{Gustavsson,
  Bylander, Yan, Forn-D\'{\i}az, Bolkhovsky, Braje, Fitch, Harrabi, Lennon,
  Miloshi et~al.}}]{RE12}
\bibinfo{author}{\bibfnamefont{S.}~\bibnamefont{Gustavsson}},
  \bibinfo{author}{\bibfnamefont{J.}~\bibnamefont{Bylander}},
  \bibinfo{author}{\bibfnamefont{F.}~\bibnamefont{Yan}},
  \bibinfo{author}{\bibfnamefont{P.}~\bibnamefont{Forn-D\'{\i}az}},
  \bibinfo{author}{\bibfnamefont{V.}~\bibnamefont{Bolkhovsky}},
  \bibinfo{author}{\bibfnamefont{D.}~\bibnamefont{Braje}},
  \bibinfo{author}{\bibfnamefont{G.}~\bibnamefont{Fitch}},
  \bibinfo{author}{\bibfnamefont{K.}~\bibnamefont{Harrabi}},
  \bibinfo{author}{\bibfnamefont{D.}~\bibnamefont{Lennon}},
  \bibinfo{author}{\bibfnamefont{J.}~\bibnamefont{Miloshi}},
  \bibnamefont{et~al.}, \bibinfo{journal}{Phys. Rev. Lett.}
  \textbf{\bibinfo{volume}{108}}, \bibinfo{pages}{170503}
  (\bibinfo{year}{2012}).

\bibitem[{\citenamefont{Fanchini et~al.}(2015)\citenamefont{Fanchini,
  Napolitano, Cakmak, and Caldeira}}]{CDD2015PRA}
\bibinfo{author}{\bibfnamefont{F.~F.} \bibnamefont{Fanchini}},
  \bibinfo{author}{\bibfnamefont{R.~d.~J.} \bibnamefont{Napolitano}},
  \bibinfo{author}{\bibfnamefont{B.}~\bibnamefont{Cakmak}}, \bibnamefont{and}
  \bibinfo{author}{\bibfnamefont{A.~O.} \bibnamefont{Caldeira}},
  \bibinfo{journal}{Phys. Rev. A} \textbf{\bibinfo{volume}{91}},
  \bibinfo{pages}{042325} (\bibinfo{year}{2015}).

\bibitem[{\citenamefont{Aiello et~al.}(2013)\citenamefont{Aiello, Hirose, and
  Cappellaro}}]{Aiello2013Composite}
\bibinfo{author}{\bibfnamefont{C.~D.} \bibnamefont{Aiello}},
  \bibinfo{author}{\bibfnamefont{M.}~\bibnamefont{Hirose}}, \bibnamefont{and}
  \bibinfo{author}{\bibfnamefont{P.}~\bibnamefont{Cappellaro}},
  \bibinfo{journal}{Nat. Commun.} \textbf{\bibinfo{volume}{4}},
  \bibinfo{pages}{273} (\bibinfo{year}{2013}).

\bibitem[{\citenamefont{L{\"u}cke et~al.}(2011)\citenamefont{L{\"u}cke,
  Scherer, Kruse, Pezz{\'e}, Deuretzbacher, Hyllus, Topic, Peise, Ertmer, Arlt
  et~al.}}]{klempt2011}
\bibinfo{author}{\bibfnamefont{B.}~\bibnamefont{L{\"u}cke}},
  \bibinfo{author}{\bibfnamefont{M.}~\bibnamefont{Scherer}},
  \bibinfo{author}{\bibfnamefont{J.}~\bibnamefont{Kruse}},
  \bibinfo{author}{\bibfnamefont{L.}~\bibnamefont{Pezz{\'e}}},
  \bibinfo{author}{\bibfnamefont{F.}~\bibnamefont{Deuretzbacher}},
  \bibinfo{author}{\bibfnamefont{P.}~\bibnamefont{Hyllus}},
  \bibinfo{author}{\bibfnamefont{O.}~\bibnamefont{Topic}},
  \bibinfo{author}{\bibfnamefont{J.}~\bibnamefont{Peise}},
  \bibinfo{author}{\bibfnamefont{W.}~\bibnamefont{Ertmer}},
  \bibinfo{author}{\bibfnamefont{J.}~\bibnamefont{Arlt}}, \bibnamefont{et~al.},
  \bibinfo{journal}{Science} \textbf{\bibinfo{volume}{334}},
  \bibinfo{pages}{773} (\bibinfo{year}{2011}).

\bibitem[{\citenamefont{Peise et~al.}(2015)\citenamefont{Peise, Kruse, Lange,
  L¨¹cke, Pezz¨¨, Arlt, Ertmer, Hammerer, Santos, and Smerzi}}]{klempt2015}
\bibinfo{author}{\bibfnamefont{J.}~\bibnamefont{Peise}},
  \bibinfo{author}{\bibfnamefont{I.}~\bibnamefont{Kruse}},
  \bibinfo{author}{\bibfnamefont{K.}~\bibnamefont{Lange}},
  \bibinfo{author}{\bibfnamefont{B.}~\bibnamefont{L¨¹cke}},
  \bibinfo{author}{\bibfnamefont{L.}~\bibnamefont{Pezz¨¨}},
  \bibinfo{author}{\bibfnamefont{J.}~\bibnamefont{Arlt}},
  \bibinfo{author}{\bibfnamefont{W.}~\bibnamefont{Ertmer}},
  \bibinfo{author}{\bibfnamefont{K.}~\bibnamefont{Hammerer}},
  \bibinfo{author}{\bibfnamefont{L.}~\bibnamefont{Santos}}, \bibnamefont{and}
  \bibinfo{author}{\bibfnamefont{A.}~\bibnamefont{Smerzi}},
  \bibinfo{journal}{Nat. Commun.} \textbf{\bibinfo{volume}{6}},
  \bibinfo{pages}{8984} (\bibinfo{year}{2015}).

\bibitem[{mag()}]{magmeter}
\bibinfo{note}{In Ref. [34], C. D. Aiello {\it et al}. experimentally
  investigate a magnetometry scheme based on rotary echos (RE) applied in a
  nitrogen-vacancy centre in diamond. After transforming the Hamiltonian to the
  toggling-frame, they obtain the evolution of the population (the signal) for
  an initial state under 55 RE cycles. Then they calculate the sensitivity and
  periodogram (the squared magnitude of the Fourier transform of the signal)
  which identifies the frequency content of the signal. By adjusting the
  rotation angle of the half-echo of RE, the coherence time and sensitivity of
  the sensor are improved in their scheme in the presence of noisy environment.
  In addition, Eto {\it et al}. demonstrate an ac magnetometry based on
  spin-echos of a spinor BEC~[11].}

\bibitem[{\citenamefont{Ho}(1998)}]{ho1998spinor}
\bibinfo{author}{\bibfnamefont{T.-L.} \bibnamefont{Ho}},
  \bibinfo{journal}{Phys. Rev. Lett.} \textbf{\bibinfo{volume}{81}},
  \bibinfo{pages}{742} (\bibinfo{year}{1998}).

\bibitem[{\citenamefont{Ohmi and Machida}(1998)}]{ohmi1998bose}
\bibinfo{author}{\bibfnamefont{T.}~\bibnamefont{Ohmi}} \bibnamefont{and}
  \bibinfo{author}{\bibfnamefont{K.}~\bibnamefont{Machida}},
  \bibinfo{journal}{J. Phys. Soc. Jpn.} \textbf{\bibinfo{volume}{67}},
  \bibinfo{pages}{1822} (\bibinfo{year}{1998}).

\bibitem[{\citenamefont{Law et~al.}(1998)\citenamefont{Law, Pu, and
  Bigelow}}]{Law98}
\bibinfo{author}{\bibfnamefont{C.~K.} \bibnamefont{Law}},
  \bibinfo{author}{\bibfnamefont{H.}~\bibnamefont{Pu}}, \bibnamefont{and}
  \bibinfo{author}{\bibfnamefont{N.~P.} \bibnamefont{Bigelow}},
  \bibinfo{journal}{Phys. Rev. Lett.} \textbf{\bibinfo{volume}{81}},
  \bibinfo{pages}{5257} (\bibinfo{year}{1998}).

\bibitem[{\citenamefont{Zhang et~al.}(2003)\citenamefont{Zhang, Yi, and
  You}}]{Zhang03}
\bibinfo{author}{\bibfnamefont{W.}~\bibnamefont{Zhang}},
  \bibinfo{author}{\bibfnamefont{S.}~\bibnamefont{Yi}}, \bibnamefont{and}
  \bibinfo{author}{\bibfnamefont{L.}~\bibnamefont{You}}, \bibinfo{journal}{New
  J. Phys.} \textbf{\bibinfo{volume}{5}}, \bibinfo{pages}{77}
  (\bibinfo{year}{2003}).

\bibitem[{\citenamefont{Yi et~al.}(2002)\citenamefont{Yi, M\"ustecapl\ifmmode
  \imath \else \i \fi{}o\ifmmode~\breve{g}\else \u{g}\fi{}lu, Sun, and
  You}}]{Yi02}
\bibinfo{author}{\bibfnamefont{S.}~\bibnamefont{Yi}},
  \bibinfo{author}{\bibfnamefont{O.~E.} \bibnamefont{M\"ustecapl\ifmmode \imath
  \else \i \fi{}o\ifmmode~\breve{g}\else \u{g}\fi{}lu}},
  \bibinfo{author}{\bibfnamefont{C.~P.} \bibnamefont{Sun}}, \bibnamefont{and}
  \bibinfo{author}{\bibfnamefont{L.}~\bibnamefont{You}},
  \bibinfo{journal}{Phys. Rev. A} \textbf{\bibinfo{volume}{66}},
  \bibinfo{pages}{011601} (\bibinfo{year}{2002}).

\bibitem[{\citenamefont{Ulrich}(1976)}]{Haeberlen76}
\bibinfo{author}{\bibfnamefont{H.}~\bibnamefont{Ulrich}},
  \emph{\bibinfo{title}{High Resolution NMR in Solids Selective Averaging}}
  (\bibinfo{publisher}{Elsevier Science}, \bibinfo{year}{1976}).

\bibitem[{\citenamefont{Zhang et~al.}(2005)\citenamefont{Zhang, Zhou, Chang,
  Chapman, and You}}]{Zhang05a}
\bibinfo{author}{\bibfnamefont{W.}~\bibnamefont{Zhang}},
  \bibinfo{author}{\bibfnamefont{D.~L.} \bibnamefont{Zhou}},
  \bibinfo{author}{\bibfnamefont{M.-S.} \bibnamefont{Chang}},
  \bibinfo{author}{\bibfnamefont{M.~S.} \bibnamefont{Chapman}},
  \bibnamefont{and} \bibinfo{author}{\bibfnamefont{L.}~\bibnamefont{You}},
  \bibinfo{journal}{Phys. Rev. A} \textbf{\bibinfo{volume}{72}},
  \bibinfo{eid}{013602} (\bibinfo{year}{2005}).

\bibitem[{\citenamefont{Ma et~al.}(2011)\citenamefont{Ma, Wang, Sun, and
  Nori}}]{ma2011quantum}
\bibinfo{author}{\bibfnamefont{J.}~\bibnamefont{Ma}},
  \bibinfo{author}{\bibfnamefont{X.}~\bibnamefont{Wang}},
  \bibinfo{author}{\bibfnamefont{C.}~\bibnamefont{Sun}}, \bibnamefont{and}
  \bibinfo{author}{\bibfnamefont{F.}~\bibnamefont{Nori}},
  \bibinfo{journal}{Phys. Rep.} \textbf{\bibinfo{volume}{509}},
  \bibinfo{pages}{89} (\bibinfo{year}{2011}).

\bibitem[{\citenamefont{Kitagawa and Ueda}(1993)}]{kitagawa1993squeezed}
\bibinfo{author}{\bibfnamefont{M.}~\bibnamefont{Kitagawa}} \bibnamefont{and}
  \bibinfo{author}{\bibfnamefont{M.}~\bibnamefont{Ueda}},
  \bibinfo{journal}{Phys. Rev. A} \textbf{\bibinfo{volume}{47}},
  \bibinfo{pages}{5138} (\bibinfo{year}{1993}).

\bibitem[{\citenamefont{Zhang et~al.}(1990)\citenamefont{Zhang, Feng, and
  Gilmore}}]{RevModPhys.62.867}
\bibinfo{author}{\bibfnamefont{W.~M.} \bibnamefont{Zhang}},
  \bibinfo{author}{\bibfnamefont{D.~H.} \bibnamefont{Feng}}, \bibnamefont{and}
  \bibinfo{author}{\bibfnamefont{R.}~\bibnamefont{Gilmore}},
  \bibinfo{journal}{Rev. Mod. Phys.} \textbf{\bibinfo{volume}{62}},
  \bibinfo{pages}{867} (\bibinfo{year}{1990}).

\bibitem[{\citenamefont{Liu et~al.}(2011)\citenamefont{Liu, Xu, Jin, and
  You}}]{liu2011spin}
\bibinfo{author}{\bibfnamefont{Y.}~\bibnamefont{Liu}},
  \bibinfo{author}{\bibfnamefont{Z.}~\bibnamefont{Xu}},
  \bibinfo{author}{\bibfnamefont{G.}~\bibnamefont{Jin}}, \bibnamefont{and}
  \bibinfo{author}{\bibfnamefont{L.}~\bibnamefont{You}},
  \bibinfo{journal}{Phys. Rev. Lett.} \textbf{\bibinfo{volume}{107}},
  \bibinfo{pages}{013601} (\bibinfo{year}{2011}).

\bibitem[{\citenamefont{Pu et~al.}(2016)\citenamefont{Pu, Zhang, Yi, Wang, and
  Zhang}}]{Pu16}
\bibinfo{author}{\bibfnamefont{Z.}~\bibnamefont{Pu}},
  \bibinfo{author}{\bibfnamefont{J.}~\bibnamefont{Zhang}},
  \bibinfo{author}{\bibfnamefont{S.}~\bibnamefont{Yi}},
  \bibinfo{author}{\bibfnamefont{D.}~\bibnamefont{Wang}}, \bibnamefont{and}
  \bibinfo{author}{\bibfnamefont{W.}~\bibnamefont{Zhang}},
  \bibinfo{journal}{Phys. Rev. A} \textbf{\bibinfo{volume}{93}},
  \bibinfo{pages}{053628} (\bibinfo{year}{2016}).

\end{thebibliography}
\end{document}